# Relational Algebra as non-Distributive Lattice

VADIM TROPASHKO
Vadim.Tropashko@orcl.com

________________________________________________________________

We reduce the set of classic relational algebra operators to two binary operations: natural join and generalized union. We further demonstrate that this set of operators is relationally complete and honors lattice axioms.

________________________________________________________________

## 1. INTRODUCTION

Chapter 4 of the Alice book ([1]) describes SPC Algebra for conjunctive queries. Selection, projection, and Cartesian product are primitive operations, while few others - intersection, for example - could be expressed as a composition of primitive ones. On the other hand, union is a notable exception, so that it is introduced as an additional primitive operation to form SPCU algebra. This apparent lack of symmetry between union and intersection has been investigated in [2], where the number of basic relational operators had been reduced to two pairs of mutually dual operators. Remarkably, set operators were excluded from the basic ones.

In this paper we reduce the set of classic relational algebra operators to two binary operations: natural join and generalized union. We demonstrate that this set of operators is relationally complete and honors lattice axioms. In the final section we draw a connection to *Formal Concept Analysis* [5].

## 2. FUNDAMENTAL OPERATORS

### 2.1 Natural Join

Natural join is formally defined as Cartesian product followed by selection with equality predicates on the columns with identical names:

$$A(x,y) \bowtie B(y,z) = \sigma_{A.y=B.y} (A(x,y) \oplus B(y,z))$$

The two identical y columns in the resulting relation are merged into a single column. As argued in [2] this action is symmetric to merging duplicate rows and is always assumed as selection side effect. In this paper we'll shift perspective and emphasize natural join as basic operator. Selection and Cartesian product would be expressed in terms of natural join.

*Example 1.* Given relations A

| x | y |
|---|---|
| 1 | 2 |
| 1 | 3 |

and B

| y | z |
|---|---|
| 1 | 3 |
| 2 | 4 |

natural join $A \bowtie B$ is

| x | y | z |
|---|---|---|
| 1 | 2 | 4 |

## 2.2 Generalized Union

Generalized union is formally defined as set union restricted to a common set of relation attributes:

$$A(x,y) \bigtimes B(y,z) = (\pi_y A(x,y)) \cup (\pi_y B(y,z))$$

Unlike traditional, set-based union, there is no requirement that relational operand signatures should match.

Alternatively, generalized union can be defined as tensor product followed by projection into a set of common columns:

$$A(x,y) \bigtimes B(y,z) = \pi_y (A(x,y) \otimes B(y,z))$$

The duplicate rows are merged as usual projection side effect in the set (as opposed to bag) semantics. This definition is dual to natural join definition of section 2.1.

*Example 2.* Given relations A and B from the example 1, generalized union $A \bigtimes B$ is

| y |
|---|
| 1 |
| 2 |
| 3 |

Perhaps, the name *generalized intersection* instead of *natural join* would better emphasize the symmetry between the two fundamental operators. We, nevertheless stick to traditional naming convention. The other possibility is calling *generalized union* as *natural* one. Even better option is omitting adjectives altogether.

## 3. RELATIONAL COMPLETENESS

In the next sections we'll examine basic operators from [2] one by one and express them in terms of natural join and generalized union.

### 3.1 Cartesian Product

Cartesian product is a natural join between relations with disjoint set of columns.

### 3.2 Selection

Selection $\sigma_{p(x,y)}A(x,y,z)$ is natural join between A and P:

$$\sigma_{p(x,y)}A(x,y,z) = A(x,y,z) \bowtie P(x,y)$$

Admittedly, the cost of this reduction is introducing potentially infinite relation corresponding to selection predicate. In author's opinion, this price is justified. Raising abstraction level and reducing complexity always produce nicer mathematics. Infinity emerges naturally in many mathematics endeavors and always simplifies underlying theory.

*Example 3a*. Given relation A(x,y), selection $\sigma_{x=1} A(x,y)$ is natural join of A with finite relation $EQ_1(x)$

| x |
|---|
| 1 |

*Example 3b*. Given relation A from the example 1, selection $\sigma_{x>1} A(x,y)$ is natural join of A with infinite relation $GT_1(x) = \{ (x) \mid x>1 \}$

| x |
|---|
| 2 |
| 3 |
| 4 |
| … |

assuming domain of integers.

*Example 3c*. Given relation $A$ from the example 1, selection $\sigma_{x<y} A(x,y)$ is natural join of $A$ with infinite relation $GT(x,y) = \{ (x,y) \mid x<y \}$

| x | y |
|---|---|
| 1 | 2 |
| 2 | 3 |
| 3 | 4 |
| 1 | 3 |
| 1 | 4 |
| 2 | 4 |
| … | … |

again, assuming domain of integers.

3.3 Projection

Projection $\pi_y A(x,y)$ is generalized union of between $A$ and empty relation $Y(y)$:

$$\pi_y A(x,y) = A(x,y) \bigotimes Y(y)$$

In order to express the remaining operator – tensor product, we have to master column renaming, first.

3.4 Renaming

In classic relational algebra renaming is somewhat obscure operation. Textbooks and online lecture notes often include it into the list of basic operations (with a dedicated symbol $\rho$), while the others treat it as an implicit manipulation upon relations. It is indeed odd operation, since the author is unaware of any other algebraic system in mathematics which has renaming operation. The answer to this problem is that renaming $\rho: A(x,y) \rightarrow A(x,z)$ could be expressed as composition of natural join between given relation $A(x,y)$ and (potentially infinite) binary identity relation $EQ(y,z) = \{ (y,z) \mid y=z \}$, and projection $\pi_{x,z}$:

$$\rho: A(x,y) \rightarrow A(x,z) = \pi_{x,z} ( A(x,y) \bowtie EQ(y,z) ) = XZ(x,z) \bigotimes ( A(x,y) \bowtie EQ(y,z) )$$

*Example 4*. Given relation $A(x,y)$, transitive closure is defined as

$$TC_A(x,y) = A \bigotimes A^2 \bigotimes A^3 \bigotimes \ldots$$

where relation powers are defined by induction with elementary step

$$A^{i+1} = A^i . A = ( \rho: A^i (x,y) \rightarrow A^i (x,s) ) \bowtie ( \rho: A (x,y) \rightarrow A (s,y) )$$

Without renaming, self join $A \bowtie A$ evaluates to $A$.

## 3.5 Difference

Reduction for difference operator in [2] involves selection, union, renaming, Cartesian product, and (natural) extension of projection operator with aggregation. Leveraging infinite relation $NEQ(x,y) = \{ (x,y) \mid x \neq y \}$ prompts alternative implementation via anti-join

$$A'(x,y) = \pi_{x,y} ( A(x,y) \bowtie ( \rho: B(x,y) \to B(x',y') ) \bowtie NEQ(x,x') )$$

$$A''(x,y) = \pi_{x,y} ( A(x,y) \bowtie ( \rho: B(x,y) \to B(x',y') ) \bowtie NEQ(y,y') )$$

$$A \setminus B = A'(x,y) \;\overline{\bowtie}\; A''(x,y)$$

which doesn't require aggregation.

At this point, we have already proved that natural join and generalized union make up a relationally complete set of operators. Nevertheless, let's demonstrate that tensor product can be expressed via natural join and generalized union, as well.

## 3.6 Tensor Product

Given relations $A(x,y)$ and $B(x,y)$ from the example 1, tensor product $A \otimes B$ can be produced in 3 steps:

$$A'(x,xx,xy,yx,yy,y) = A(x,y) \bowtie EQ(x,xx) \bowtie EQ(x,xy) \bowtie EQ(y,yx) \bowtie EQ(y,yy)$$
$$B'(x,xx,xy,yx,yy,y) = A(x,y) \bowtie EQ(x,xx) \bowtie EQ(y,xy) \bowtie EQ(x,yx) \bowtie EQ(y,yy)$$

$$A \otimes B = \pi_{xx,xy,yx,yy} ( A'(x,xx,xy,yx,yy,y) \;\overline{\bowtie}\; B'(x,xx,xy,yx,yy,y) )$$

## 4. LATTICE THEORY PERSPECTIVE

Algebra with two binary operations $\wedge$ and $\vee$ is called *lattice* [4] if those operations are idempotent, commutative, associative, and satisfy the absorption law. Lattice operations $\wedge$ and $\vee$ are called *meet* and *join*, correspondingly.

There are 2 possibilities to match lattice operators against natural join and generalized union, but associating lattice join with natural join, perhaps, would be the best to avoid confusion. We write lattice axioms with relational operations as follows:

## Idempotent laws:

$A \bowtie A = A \qquad\qquad A \;\overline{\bowtie}\; A = A$

**Commutative laws:**

$A \bowtie B = B \bowtie A$  $\qquad A \owtie B = B \owtie A$

**Associative laws:**

$A \bowtie (B \bowtie C) = (A \bowtie B) \bowtie C$  $\qquad A \owtie (B \owtie C) = (A \owtie B) \owtie C$

**Absorption laws:**

$A \owtie (A \bowtie B) = A$  $\qquad A \bowtie (A \owtie B) = A$

The first three laws for natural join are well known. Their dual counterparts easily follow from union and intersection being idempotent, commutative and associative (union being applied to columns, and intersection to rows).

Proving absorption law just a little bit more involved. For any relation $A'$ with signature containing signature of $A$ the following condition is satisfied

$$A(x,y) \subseteq A(x,y) \owtie A'(x,y,z)$$

Therefore,

$$A \subseteq A \owtie (A \bowtie B)$$

Conversely, both

$$A(x,y) \owtie (A(x,y) \bowtie B(y,z)) = A(x,y) \cup \pi_{x,y}(A(x,y) \bowtie B(y,z))$$

and

$$\pi_{x,y}(A(x,y) \bowtie B(y,z)) \subseteq A(x,y)$$

imply

$$A \owtie (A \bowtie B) \subseteq A$$

4.1 Partial order

From the lattice we obtain partially ordered set by defining

$$A \leq B \quad \text{iff} \quad B = A \bowtie B$$

Symmetrically,

$$A \leq B \quad \text{iff} \quad A = A \owtie B$$

## 4.2 Least and Greatest Elements

Let's meet all lattice elements and denote the resulting relation as $\mathrm{DUM}$. Symmetrically let denote join of all relations as $\mathrm{DEE}$. Then, for any relation $A$

$$\mathrm{DUM} \leq A \leq \mathrm{DEE}$$

and, therefore

$$A \bowtie \mathrm{DUM} = A$$
$$A \veebar \mathrm{DUM} = \mathrm{DUM}$$
$$A \bowtie \mathrm{DEE} = \mathrm{DEE}$$
$$A \veebar \mathrm{DEE} = A$$

Clearly, generalized union of all relations implies that

1. intersection of all relation signatures must be empty, and
2. union of all possible rows must have at least one row.

Hence, the relation with no columns, one row and the symbol $\mathrm{DUM}$. (Set semantics allows only one row in an empty signature relation).

Likewise, joining all the relations would produce relation with no rows $\mathrm{DEE}$.

In [3] these relations are known as DUM and DEE.

## 4.3 Distributive and Modular Properties

Unlike set intersection and union, natural join and generalized union don't honor distributive law. Otherwise, if relational lattice were distributive it would be isomorphic to some lattice of sets. Our lattice, however, is a *cylindrical sets* algebra: while the join of two cylindrical sets is always the same as their set intersection, the meet of two cylindrical sets defined as their set union wouldn't produce a cylindrical set as a result.

*Example 5*. Given the relations $A$ and $B$ from the example 1, and the relation $C$

| z |
|---|
| 3 |
| 7 |

the reader can verify

$$(A \veebar B) \bowtie C \neq (A \veebar C) \bowtie (B \veebar C)$$

The lattice in the example 5 contains a sublattice isomorphic to $N_5$. By $M_3$-$N_5$ theorem [4] it follows that it doesn't satisfy *modular* property either.

## 5. Case Study: Formal Concepts Analysis

*Formal Concepts Analysis* [5] is a method for data analysis, knowledge representation and information management based on lattice theory foundation. Unlike relational theory, this niche discipline is widely unknown in the US. On quick examination it turns out that the lattice operations in Formal Concept Analysis are identical to the ones introduced in this paper.

*Example 6* ([5]). A formal context of "famous animals"

|          | cartoon | real | tortoise | dog | cat | mammal |
|----------|---------|------|----------|-----|-----|--------|
| Garfield | x       |      |          |     | x   | x      |
| Snoopy   | x       |      |          | x   |     | x      |
| Socks    |         | x    |          |     | x   | x      |
| Bobby    |         | x    |          | x   |     | x      |
| Harriet  |         | x    | x        |     |     |        |

produces a concept lattice with the following Hasse diagram

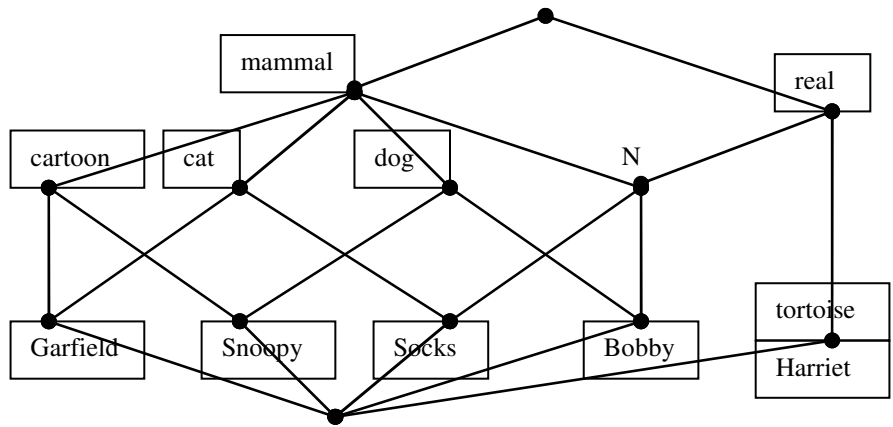

The lattice elements can be interpreted as relations. Each relation must have at least one column Name. For each *formal attribute* element, which is greater than a node in question, an additional column is introduced. In the example 6 the relation corresponding to the lattice element N has the signature {Name, isMammal, isReal}. For each *object* element, which is less than the node, the relation would contain a row. All the fields are assigned the same value true, except the Name, which is the object name. The lattice element N in the example 6 is the relation

| Name | isMammal | isReal |
|------|----------|--------|
| Socks | true | true |
| Bobby | true | true |

Admittedly, the produced relations are rather special. If nothing else, the connection between relational theory and formal concepts analysis allows exchange of methods between the both fields.